\documentclass[preprint]{aastex}
\usepackage{rotating}

\def\cs{c_{\rm s}}

\def\K{\,{}\mbox{K}}

\def\Mach{\mathcal{\,M}}
\def\kms{\mbox{\,km}\mbox{\,s}^{-1}}

\def\Lbox{L_{\rm box}}
\def\LJ{L_{\rm J}}

\def\Mb{M_{\rm box}} 
\def\MBE{M_{\rm BE}}
\def\Mcore{M_{\rm core}}

\def\Myr{\mbox{\,Myr}}
\def\Msun{\mbox{M}_{\odot}}
\def\pc{\mbox{\,pc}}

\def\pcc{\mbox{\,cm}^{-3}}
\def\rf{r_{\rm f}}
\def\rhob{\rho_{\rm b}}
\def\rhoc{\rho_{\rm c}}

\def\tff{t_{\rm{ff}}}

\def\eg{{e.g.}}
\def\x{\times}

\def\icms{\mbox{\,cm}^{-2}}
\def\icmc{\mbox{\,cm}^{-3}}

\def\H2{\mbox{H}_2}                  

\def\part#1#2{{\frac{\partial{#1}}{\partial{#2}}}}

\slugcomment{Submitted to ApJ}

\shorttitle{Collapsing cores within collapsing clouds}
\shortauthors{Naranjo-Romero, V\'azquez-Semadeni \& Loughnane}

\begin{document}

\title{Hierarchical gravitational fragmentation. I. Collapsing cores
  within collapsing clouds}

\author{Ra\'ul Naranjo-Romero\altaffilmark{1}, Enrique
  V\'azquez-Semadeni\altaffilmark{1}, and Robert M.\
  Loughnane\altaffilmark{1}} 

\altaffiltext{1}{Instituto de Radioastronom\'\i a y Astrof\'\i sica,
Universidad Nacional Aut\'onoma de M\'exico, Apdo. Postal 3-72, Morelia,
Michoac\'an, 58089, M\'exico}

\begin{abstract}
  We investigate the Hierarchical Gravitational Fragmentation scenario through numerical simulations of the prestellar stages of the collapse of a marginally gravitationally unstable isothermal sphere immersed in a strongly gravitationally unstable, uniform background medium. The core developes a Bonnor-Ebert (BE)-like density profile, while at the time of singularity (the protostar) formation the envelope approaches a singular-isothermal-sphere (SIS)-like $r^{-2}$ density profile. However, these structures are never hydrostatic. In this case, the central flat region is characterized by an infall speed, while the envelope is characterized by a uniform speed. This implies that the hydrostatic SIS initial condition leading to Shu's classical inside-out solution is not expected to occur, and therefore neither should the inside-out solution. Instead, the solution collapses from the outside-in, naturally explaining the observation of extended infall velocities. The core, defined by  the radius at which it merges with the background, has a time-variable mass, and evolves along the locus of the ensemble of observed prestellar cores in a plot of $M/\MBE$ {\it vs.} $M$, where $M$ is the core's mass and $\MBE$ is the critical Bonnor-Ebert mass, spanning the range from the ``stable'' to the ``unstable'' regimes, even though it is collapsing at all times. We conclude that the presence of an unstable background allows a core to evolve dynamically from the time when it first appears, even when it resembles a pressure-confined, stable BE-sphere. The core can be thought of as a ram-pressure confined BE-sphere, with an increasing mass due to the accretion from the unstable background.
\end{abstract}


\section{Introduction}

\subsection{The scenario of hierarchical gravitational collapse}
\label{sec:introduction}

Molecular clouds (MCs) are associated with the bulk of star formation in
the Galaxy. With supersonic linewidths \citep{Wilson70} that scale as
R$^{1/2}$, where R is the cloud's radius \citep{Larson81, Solomon87}, and
low star formation efficiencies \citep{Myers86}. MCs have traditionally
been interpreted as being in a global state of virial equilibrium
between supersonic turbulence and self-gravity. On the smallest scales,
when the density increases, the turbulence is believed to dissipate,
allowing collapse to proceed \citep{Goodman98, Tafalla02, Pineda10}.

However, recent numerical studies suggest that the picture of gravitational
collapse is not limited to the scale of dense cores, but instead may extend
to the scale of the whole cloud \citep[\eg] [] {VS+07, VS+09, VS+10, VS+11, HH08, Heitsch+08}. Moreover, several observational studies have shown that gravitational collapse extends at least to parsec scales 
\citep[\eg][]{Galvan09, Schneider10, Polychroni13, Peretto+13}.
Finally, \citet{BP11a}, extending the data presented by \citet{Heyer09},
have shown that the energy budget of giant molecular clouds (GMCs) and
high-mass star-forming clumps alike is consistent with generalized
free-fall in these structures. In this case, what has been previously
interpreted as virialization, may just as well be interpreted as
free-fall, within the uncertainties, and in fact, the data are
marginally more consistent with free-fall than with virial equilibrium.

In addition, recent observations \citep[\eg][]{Gutermuth08, Myers09,
 Andre10, Menshchikov10, Molinari10, Arzoumanian11} have revealed a
vast network of filaments everywhere inside the MCs, feeding the clumps
and the dense cores. Numerical simulations of cloud formation
\citep[\eg][]{BurkertHartmann04, HartmannBurkert07, Heitsch09, VS+07,
 VS+09, VS+11, GV14} also exhibit such filamentary structure, and
\citet{GV14} and \citet{Smith+14} have interpreted it as the consequence
of anisotropic, large-scale gravitational collapse in the clouds.
Observations of the kinematics in the filaments \citep{Schneider10,
 Kirk13, Peretto+14} are consistent with the kinematics seen in those
simulations. 

In the scenario of global hierarchical gravitational collapse and
fragmentation, then, the entire cloud is gravitationally collapsing.
This is possible because of the coherent production of cold atomic gas
over large scales by large-scale shock-compressed layers in the warm,
diffuse medium, which trigger thermal instability and a phase transition
to the cold phase \citep{BP+99, HP99, KI00, KI02, AH05, Heitsch05, VS+06}. 
The cold cloud formed by this mechanism is supersonically
turbulent, albeit, only moderately.  Within these large-scale unstable
clouds, small-scale, nonlinear density fluctuations produced by
turbulence terminate their collapse before the cloud at large does,
because their free-fall times are shorter \citep{HH08}. This implies
that the clumps and cores constitute smaller-scale collapse events embedded
within larger-scale ones (at the whole cloud scale), amounting to a
hierarchical state of collapse that is moreover chaotic (with multiple
collapse centers and irregular geometries). Thus, in this paper we
investigate the collapse of an unstable core embedded in a larger, also
unstable cloud.

\subsection{Core collapse theory}
\label{sec:dense_cores}

Within this scenario of global, multi-scale collapse, it is pertinent to
re-examine some classical and recent works on the collapse of dense
cores. The seminal works of \citet{Larson69} and \citet{Penston69}, to
which we collectively refer to as LP, independently found solutions in
which collapse proceeds outside-in, developing an $r^{-2}$ density
profile in the core envelope and a velocity profile that approaches a
constant value as the core accretes mass from its envelope.
\citet{Larson69} considered uniform initial densities, while
\citet{Penston69} considered slightly centrally-condensed density
configurations. Thus, they considered the collapse starting from times
earlier than the development of a singularity (the protostar).

\citet [hereafter S77] {Shu77} studied this problem analytically,
adopting a singular isothermal sphere (SIS) as the initial {\it
 equilibrium} condition. This implies that he only considered the
evolution starting from the time of protostar formation. He assumed that
the SIS is somehow destabilized at $t=0$, causing the innermost regions
to begin collapsing, and producing a rarefaction wave that propagates
outwards, leaving a free-falling ($\rho \propto r^{-3/2}$, $v \propto
r^{-1/2}$) region behind it, while the region outside the front remains
static.  This is the well-known ``inside-out'' collapse solution, which
is generally assumed to represent the velocity field inside the cores.

Some time later, \citet[hereafter WS85]{WhitworthSummers85} investigated
the parameter space of the initial conditions, finding that the latter
can be parametrized by the initial density and the mass at the centre of
the collapsing cloud. WS85 broadly divided these solutions in three
bands (see Fig.\ 2 of that paper), considering, in general, the
contraction from times earlier than the formation of the singularity.
In particular, they noted that, if the collapsing core starts out very
far from equilibrium, it starts to contract immediately at all radii. In
this case, the overall evolutionary pattern consists of a compression
wave that starts far from the core's center and propagates {\it
 inwards}. This wave front divides the core into an inner ($r < \rf$,
where $\rf$ is the instantaneous radial position of the front) and an
outer ($r > \rf$) region. The transition between the two regions is
smooth, in spite of the fact that eventually supersonic speeds develop
in the outer region. The density profile in the inner region is nearly
uniform, while outside of the front it decays as $r^{-2}$. The velocity
profile is linear with radius in the inner region, and uniform in the
outer region.  When the wave front reaches the center, a point of
finite mass is formed at the center (the protostar), and a rarefaction
wave propagates outward, leaving behind free-fall density and velocity
profiles.

Using numerical simulations, \citet[hereafter FC93]{FosterChevalier93} investigated the
collapse of thermal-pressure bounded spheres near hydrostatic
equilibrium, finding results consistent with the LP solution, and
confirming that supersonic velocities develop in the outer region, and
eventually reach the center. More recently, \citet{Simpson11} have
suggested that prestellar cores accrete quasi-statically until they
reach their Jeans mass, and from that point onwards the cores collapse
dynamically. However, if the core is accreting it is hard to understand
why it would be hydrostatic out to a certain radius in the first
place. \citet[] [hereafter MS13; see also Vorobyov \& Basu 2005]
{Mohammadpour13} have explored a new set of 
boundary conditions for the problem, using inflow boundaries, across
which gas enters subsonically, representing accretion onto the core. They
again found that supersonic velocities develop near the time of
protostar formation, and concluded that such a setup may not be
realistic, and that magnetic support may be necessary in order to
prevent supersonic infall velocities at the time of protostar formation,
as seemingly required by observations that low-mass cores
generally have subsonic infall velocities \citep [e.g.,
] [] {Lee+99, Lee+01, Tafalla+04, Pineda10}.

In addition, there is the well-known ``luminosity problem''
\citep[e.g., ] [] {Kenyon+90} for Class 0 and Class I sources, namely
that their low luminosities would imply low accretion rates, and
therefore low infall speeds, if the material were to accrete directly
onto the protostellar surface. However, a number of authors \citep
[e.g., ] [] {KH95, WW01, DV12} have argued that this problem can be
solved if the accretion of material does not occur directly onto the
protostar, but first onto a disk and from there onto the protostar, and in
an episodic rather than uniform manner. Therefore, in this paper we
shall not be concerned with the luminosity problem.

\subsection{Starless and prestellar cores}
\label{sec:prestellar_protostellar_cores}

Traditionally, dense cores within MCs have been classified as belonging
to either of two main sub-classes, namely starless and protostellar,
depending on whether they lack or contain protostellar objects,
respectively. The ``starless'' class includes both cores that may never
form stars as well cores that will eventually do so; the latter are
called ``prestellar'', and operationally defined as a gravitationally
bound starless core \citep [e.g., ] [] {Andre+14}.  Starless cores that
appear unbound are thought to require external pressure confinement
\citep [e.g., ] [] {BM92, Lada08}.  Interestingly, \citet{Foster09} found
that almost all cores in Perseus, even prestellar cores, seem to be
gravitationally bound (see Fig.\ 11 in their paper), although it should
be noted that Perseus cores are more massive than those in the Pipe
Nebula, which appear pressure-confined \citep[\eg,][]{Lada08}. As
already mentioned in Sec.\ \ref{sec:dense_cores}, prestellar cores
exhibit infall profiles suggesting subsonic infall velocities \citep
{Lee+99, Lee+01, Tafalla+04}, although \citet{Lee+01} noted that the
inward motions are too extended to be consistent with the inside-out
collapse model of S77.

Observations \citep[e.g., ] [] {Alves01, Lada07} have shown that
prestellar cores have a Bonnor-Ebert (BE)-like \citep{Ebert55, Ebert57,
 Bonnor56} density profile which is nearly flat in the innermost
region, while at larger radii resembles a SIS density profile ($\rho \sim
r^{-2}$). It is worth noting that these profiles represent,
respectively, the regular and singular solutions of the Lane-Emden
equation of \textit{hydrostatic} balance, although there have been
suggestions that cores formed by strong turbulent compressions can also
evolve along a sequence of BE-like density profiles \citep{BP+03,
 Gomez07, GO09}. On the other hand, \citet{Keto+15} have concluded, by
comparing synthetic line profiles of several different models of
gravitational collapse with observations of the L1544 core, that only
the quasi-equilibrium contraction of an unstable BE-sphere \citep {KC10}
is consistent with the observations. However, in this case, the initial
BE-like profile is an assumed initial condition, rather than an outcome
of the simulations.

\subsection{This work}
\label{sec:this_work}

In the present paper, we investigate a more unified scenario of the
gravitational collapse of cores based on the notion that MCs are
collapsing as a whole, and that cores form, grow and collapse within
this globally collapsing environment. We start with a generic
low-amplitude gaussian density fluctuation, to which we refer to as ``the
core'', embedded within an unstable uniform density distribution, to
which we will refer to as ``the cloud'', allowing the core to develop its
density and velocity profiles self-consistently as it grows within the
medium, thus relaxing any initial assumptions about those profiles. This
study can be considered an extension of the work of MS13, by allowing
the presence of a large envelope of uniform density, which
represents the ``background'' often present in observations of cores
\citep[e.g., ] [] {Andre+14}.

The paper is structured as follows: In \S\ref{sec:simulations} we
describe the numerical simulation. The results from the simulations are
described in \S \ref{sec:results}. In \S
\ref{sec:comp_obs} we compare the results of the
simulation with observations, in particular in terms of the locus of
observed cores in a diagram describing core stability, while in \S
\ref{sec:comp_num_analy}, we discuss the evolution of the dense cores in
the context of earlier analytical solutions. Next, in \S
\ref{sec:implic_obs_samp} we discuss some implications of our
results. Finally, in \S \ref{sec:conclusions} we present a summary and
some conclusions.

\section{The simulations}
\label{sec:simulations}

We have performed numerical simulations of the collapse of a spherically
symmetric clump inside a collapsing cloud, using a spectral, fixed mesh
numerical code \citep{Leorat90, VS+10}. Because sink particles have not
been implemented into this code, and because it cannot follow very large
gradients in the variables, we limit our study to the prestellar stage
of the evolution of the dense core collapse. However, this is sufficient
for investigating the development of the initial conditions for star
formation.

We consider an isothermal gas with a mean density of $\langle 
n\rangle=10^{4} \icmc$, a mean particle weight of 2.36, and a kinetic temperature 
of T=$11.4\K$, implying an isothermal sound speed $c_{\rm{s}}= 0.2 \kms$ in a 
numerical box with periodic boundaries. The gas is initially at rest
and no gravity-counteracting forces such as a magnetic field or
small-scale turbulence are included, so the gas is strongly
Jeans-unstable. We choose the box side as $\Lbox = \sqrt{10} \, \LJ
\approx 3.16 \, \LJ \approx 0.71$ pc, where $\LJ \approx 0.22 \pc$ is
the Jeans length. The box mass is $\Mb \approx 206\, \Msun$.

The initial density field consists of a uniform background with $n
\approx 10^4 \pcc$, on
top of which we have added a density fluctuation with a Gaussian
profile, whose peak is at the box center and has a density $n \approx 1.5
\times 10^4 \icmc$ (or $\sim 50\%$ above the mean) and a FWHM
$\approx0.06\pc$. Because the fluctuation is small in size ($r_{\rm core} 
\sim 0.14 \pc$) and mass ($m_{\rm core} \approx 7.35 \, \Msun$), the background 
density is almost the same as the mean density, which is exactly $10^4 \pcc$. 
The free-fall time for the background density is $\tff=
\sqrt{3\pi/32G\rho}\approx 0.34 \Myr$. The evolution proceeds on a
timescale longer than $\tff$ because the initial gradient of the
gravitational potential is very mild and of the periodic boundary
conditions. Our initial setup is within 'band $0$' of WS85, which
corresponds to clouds that are initially centrally peaked, far from
hydrostatic equilibrium and collapsing immediately (the LP solutions
also fall within this band, as noted by WS85).

We have also tested different sets of parameters for the background
density and the central density fluctuation, but the behavior of these
collapsing structures is qualitatively the same, and thus we focus only
on the simulation described above. The only exception occurs in the
cases when the central peak contains less than the local Jeans mass. In
these cases, the density peak first expands and then collapses once
enough mass has been accreted at the center.

We have chosen this setup inspired by our observation of how
fragmentation proceeds in numerical simulations of cloud formation and
evolution \citep[e.g.,] [] {VS+07, VS+09, VS+10, HH08, Banerjee+09,
Colin+13, GV14}. It has been argued in these papers that a forming
GMC (and its atomic precursor) may rapidly acquire a large number of
Jeans masses if the converging flow that assembles it is coherent and
extended over a large region, as may be expected, for example, in a
spiral arm, or in the collect-and-collapse scenario at the border of
expanding shells \citep{ElmegreenLada97}. In this case,
\citet{Hoyle53}-type fragmentation may be expected, where successively
smaller scales may go unstable as the mean cloud density increases and
the average Jeans mass decreases due to the global collapse in a
nearly isothermal medium. The small-scale density fluctuations located
far from the trough of the large-scale potential well may start
growing locally while simultaneously being transported by the
large-scale flow towards the remote collapse center, in a ``conveyor
belt'' mode \citep[see e.g.,] [] {Longmore+14, GV14}.  As a first
approximation, this type of flow can be represented locally as a
static unstable background with a small clump collapsing within it,
because the large-scale flow towards the distant global collapse
center locally appears as a uniform bulk background motion, which is
dynamically irrelevant to the local collapse.

\section{Results}
\label{sec:results}

\begin{figure}
\includegraphics[scale=0.37, keepaspectratio]{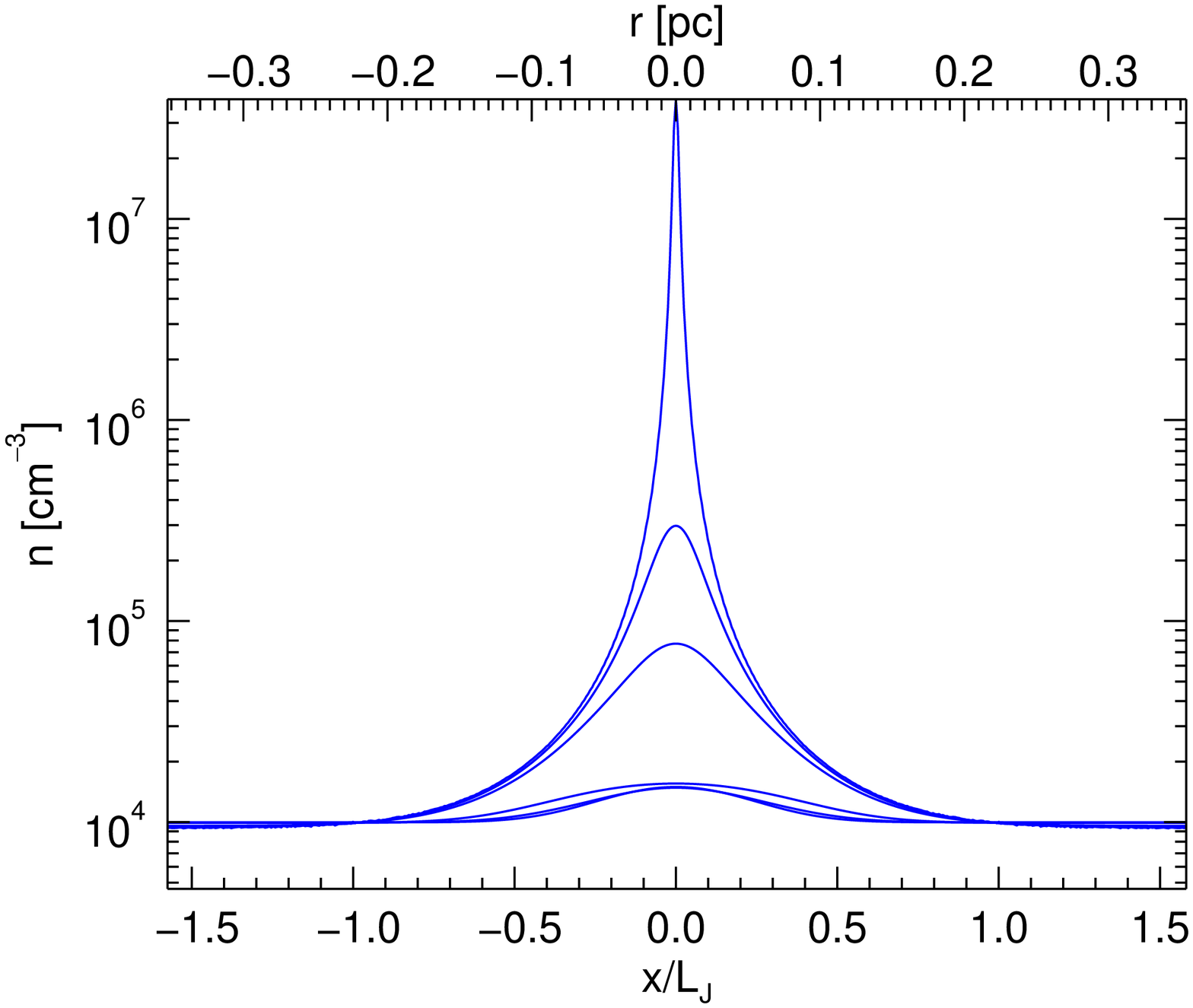}
\includegraphics[scale=0.37, keepaspectratio]{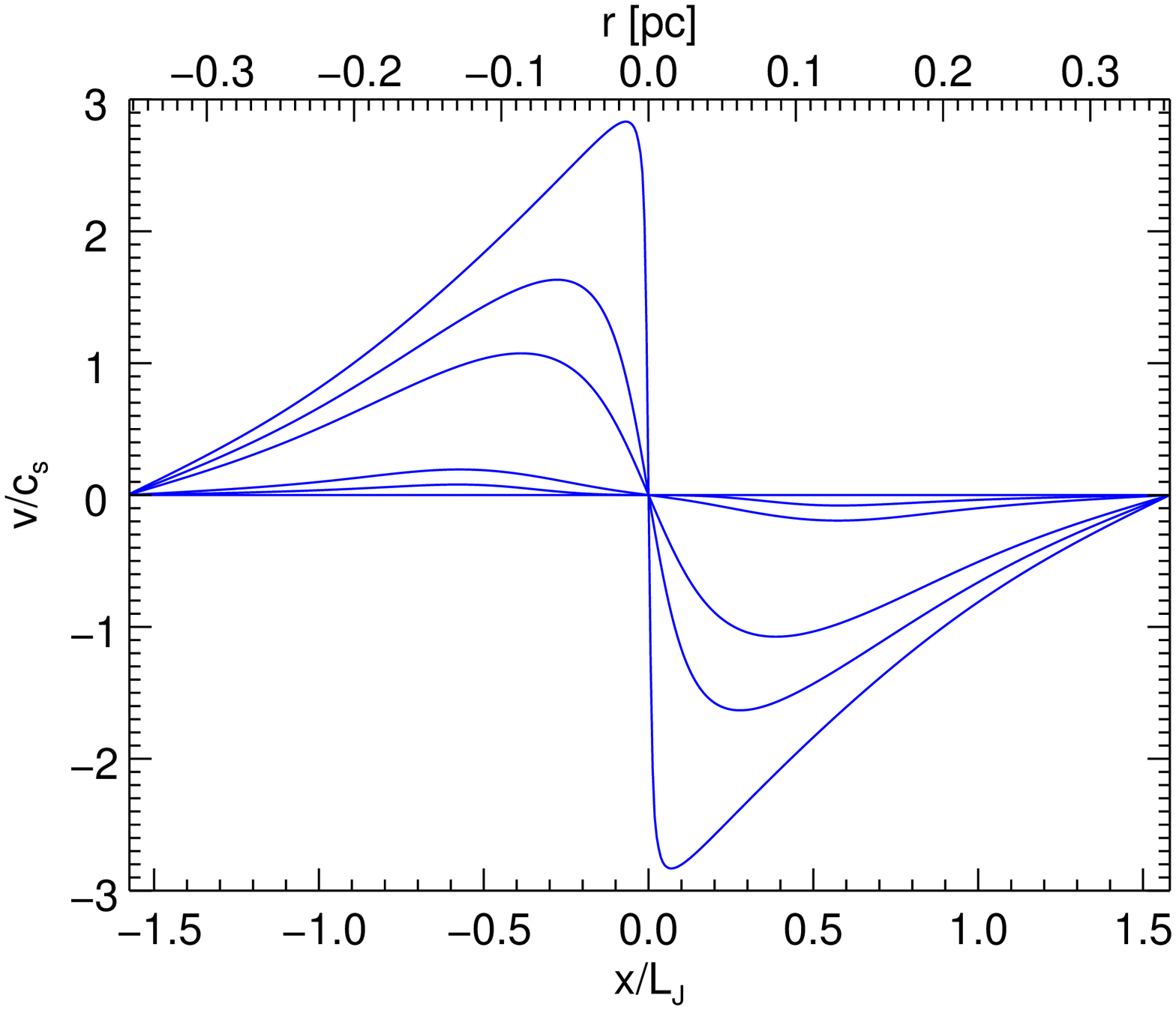}
\caption{Evolution of the density (top panel) and velocity (bottom
  panel) profiles of the simulated core (See text). The various
  lines in both panels represent the timesteps listed in Table
  \ref{tab:core_parameters}.\label{fig:dens_vel_evol}}
\end{figure}

In Fig.\ \ref{fig:dens_vel_evol}, we show various snapshots from the
evolution of the radial profiles of the density ({\it left panel}) and
the velocity ({\it right panel}) of the core and its environment, both
with a linear radial axis, to emphasize the external structure of the
core and its envelope.
It can be seen from this figure that, by embedding the
core in a uniform background medium that is also collapsing, the nature
of the collapse deviates from the self-similar asymptotic structure of
the prestellar (or pre-singularity) stages of the collapse, such as the
LP solutions. The core has a natural, well-defined ``boundary'', namely
the place where it merges into the background. This boundary increases
in radius during the evolution, but the material outside the boundary
remains at the background density, and simply develops an inflow
velocity that is continuous across the boundary, steadily accreting onto
the core. On the other hand, in agreement with previous studies
\citep{Larson69, Penston69, FosterChevalier93, Mohammadpour13}, it can
also be seen from Fig.\ \ref{fig:dens_vel_evol} that the flow remains
subsonic during most of the prestellar evolution, although
it eventually becomes supersonic at a so-called {\it sonic point}.
Subsequently, the sonic point splits into two such points, to which we
refer to as the inner  and outer sonic points. The inner sonic point
approaches the core center as time progresses, and finally reaches it at
the time of singularity formation.

\begin{table}

\begin{center}
\caption{Physical properties of the simulated core.\label{tab:core_parameters}}
\begin{tabular}{crrrrrrrrrrr}
\tableline
\tableline
\multicolumn{2}{c}{Time} & \multicolumn{1}{c}{$n_{\rm{core}}/n_{\rm{cloud}}$} & \multicolumn{1}{c}{$N_{\rm{core}}/N_{\rm{cloud}}$} & \multicolumn{1}{c}{$v_{\rm{max}}/c_{\rm{s}}$} & \multicolumn{1}{c}{$r_{\mu=1}$} & \multicolumn{1}{c}{$M_{\mu=1}$} & \multicolumn{1}{c}{${v/c_{\rm{s}}},_{\mu=1}$} \\
\multicolumn{1}{c}{$(t_{\rm{ff}})$} & \multicolumn{1}{c}{$(\Myr)$} & \multicolumn{1}{c}{\ } & \multicolumn{1}{c}{\ } & \multicolumn{1}{c}{\ } & \multicolumn{1}{c}{$(\pc)$} & \multicolumn{1}{c}{$(\Msun)$}\\
\tableline
0.00 & 0.00 & 1.50 & \multicolumn{1}{r}{1.09} & \multicolumn{1}{r}{0.00} & \multicolumn{1}{r}{0.133} & \multicolumn{1}{r}{6.39} & \multicolumn{1}{r}{0.00}\\
0.46 & 0.16 & 1.49 & \multicolumn{1}{r}{1.10} & \multicolumn{1}{r}{0.08} & \multicolumn{1}{r}{0.130} & \multicolumn{1}{r}{6.18} & \multicolumn{1}{r}{0.08}\\
0.92 & 0.31 & 1.57 & \multicolumn{1}{r}{1.14} & \multicolumn{1}{r}{0.19} & \multicolumn{1}{r}{0.124} & \multicolumn{1}{r}{5.98} & \multicolumn{1}{r}{0.19}\\
1.84 & 0.62 & 8.05 & \multicolumn{1}{r}{2.10} & \multicolumn{1}{r}{1.07} & \multicolumn{1}{r}{0.072} & \multicolumn{1}{r}{3.47} & \multicolumn{1}{r}{1.06}\\
2.01 & 0.67 & 31.43 & \multicolumn{1}{r}{3.61} & \multicolumn{1}{r}{1.63} & \multicolumn{1}{r}{0.043} & \multicolumn{1}{r}{2.09} & \multicolumn{1}{r}{1.56}\\
2.14 & 0.72 & 4224.50 & \multicolumn{1}{r}{36.10} & \multicolumn{1}{r}{2.83} & \multicolumn{1}{r}{0.003} & \multicolumn{1}{r}{0.19} & \multicolumn{1}{r}{2.07}
\end{tabular}
\tablecomments{The rows correspond to same snapshots as in fig.~\ref{fig:all_profiles}}
\end{center}
\end{table}

\begin{sidewaystable}
\centering
\caption{Physical properties of the simulated core at the different background density thresholds.\label{tab:core_parameters_sigma}}
\begin{tabular}{cccccccccccccc}
\tableline
\tableline
\multicolumn{2}{c|}{} & \multicolumn{4}{c|}{$1.125n_{\rm{bg}}$} & \multicolumn{4}{c|}{$1.250n_{\rm{bg}}$} & \multicolumn{4}{c|}{$1.50n_{\rm{bg}}$} \\ \tableline
\multicolumn{2}{c|}{Time} &$R$ & $M$ & $M_{\rm{BE}}$ & \multicolumn{1}{c|}{$\langle n\rangle$} & $R$ & $M$ & $M_{\rm{BE}}$ & \multicolumn{1}{c|}{$\langle n\rangle$} & $R$ & $M$ & $M_{\rm{BE}}$ & $\langle n\rangle$\\ 
\multicolumn{1}{c}{($t_{\rm{ff}}$)} & \multicolumn{1}{c|}{($\Myr$)} & ($\pc$) & ($\Msun$) & ($\Msun$) &\multicolumn{1}{c|}{($10^{4}\icmc$)} & ($\pc$) & ($\Msun$) & ($\Msun$) & \multicolumn{1}{c|}{($10^{4}\icmc$)} & ($\pc$) & ($\Msun$) & ($\Msun$) & ($10^{4}\icmc$) \\
\tableline
\multicolumn{1}{c}{0.00} & \multicolumn{1}{c|}{0.00} & 0.74 & 1.73 & 1.99 & \multicolumn{1}{c|}{1.23} & 0.52 & 0.66 & 1.92 & \multicolumn{1}{c|}{1.33} &  &  &  & \\
\multicolumn{1}{c}{0.69} & \multicolumn{1}{c|}{0.73} &  &  &  & \multicolumn{1}{c|}{ } &  & & & \multicolumn{1}{c|}{} & 0.05 & $5.74\x10^{-4}$ & 1.81 & 1.49\\
\multicolumn{1}{c}{2.14} & \multicolumn{1}{c|}{2.27} & 1.93 & 35.44 & 1.77 & \multicolumn{1}{c|}{1.57} & 1.57 & 25.26 & 1.58 & \multicolumn{1}{c|}{1.97} & 1.27 & 17.99 & 1.36 & 2.64\\
\end{tabular}
\tablecomments{Only the initial and final derived values are shown. The snapshots corresponds to the same snapshots as in fig.~\ref{fig:all_profiles}}
\end{sidewaystable}

\begin{figure*}[pt]
\includegraphics[scale=.38]{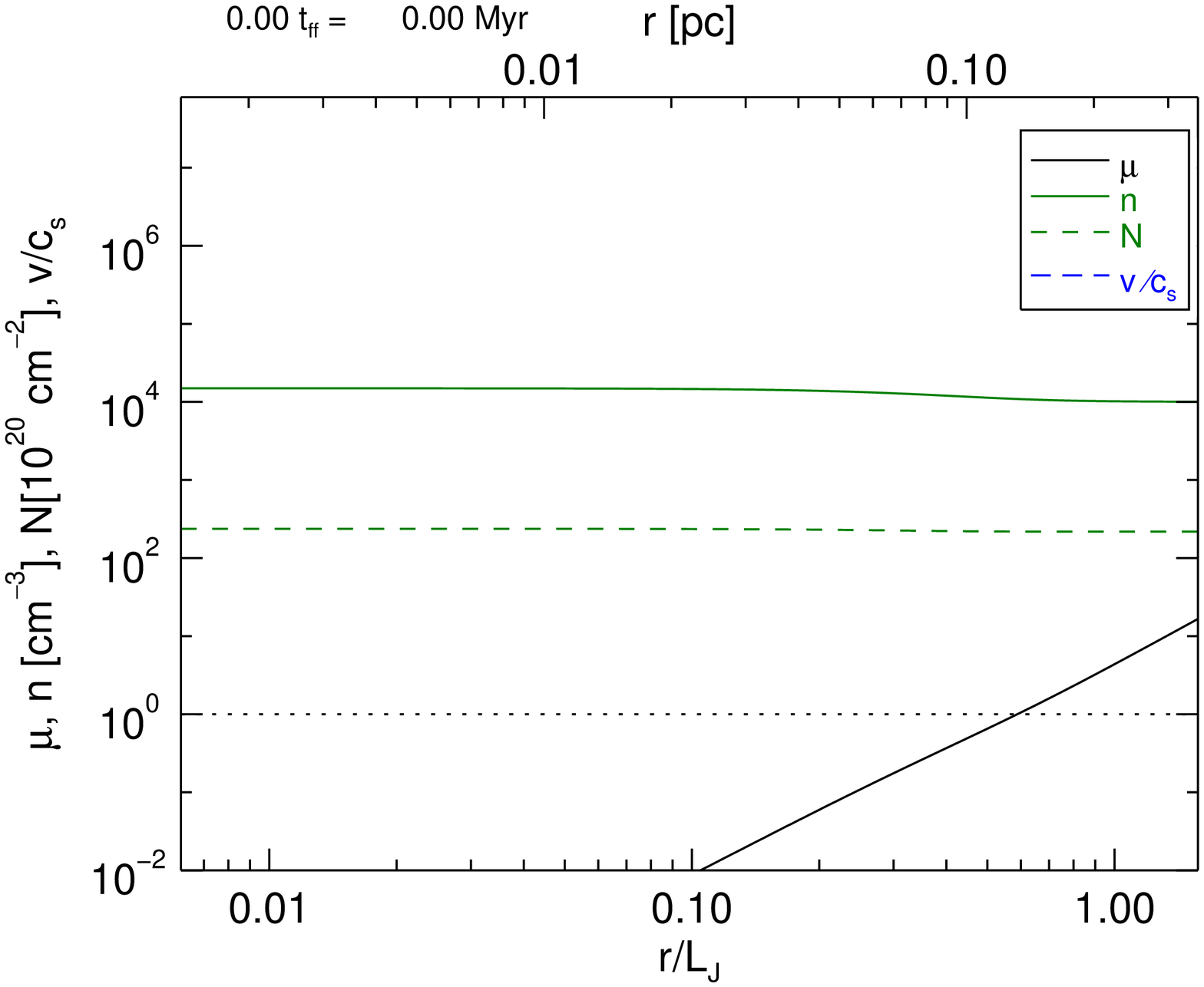}\includegraphics[scale=.38]{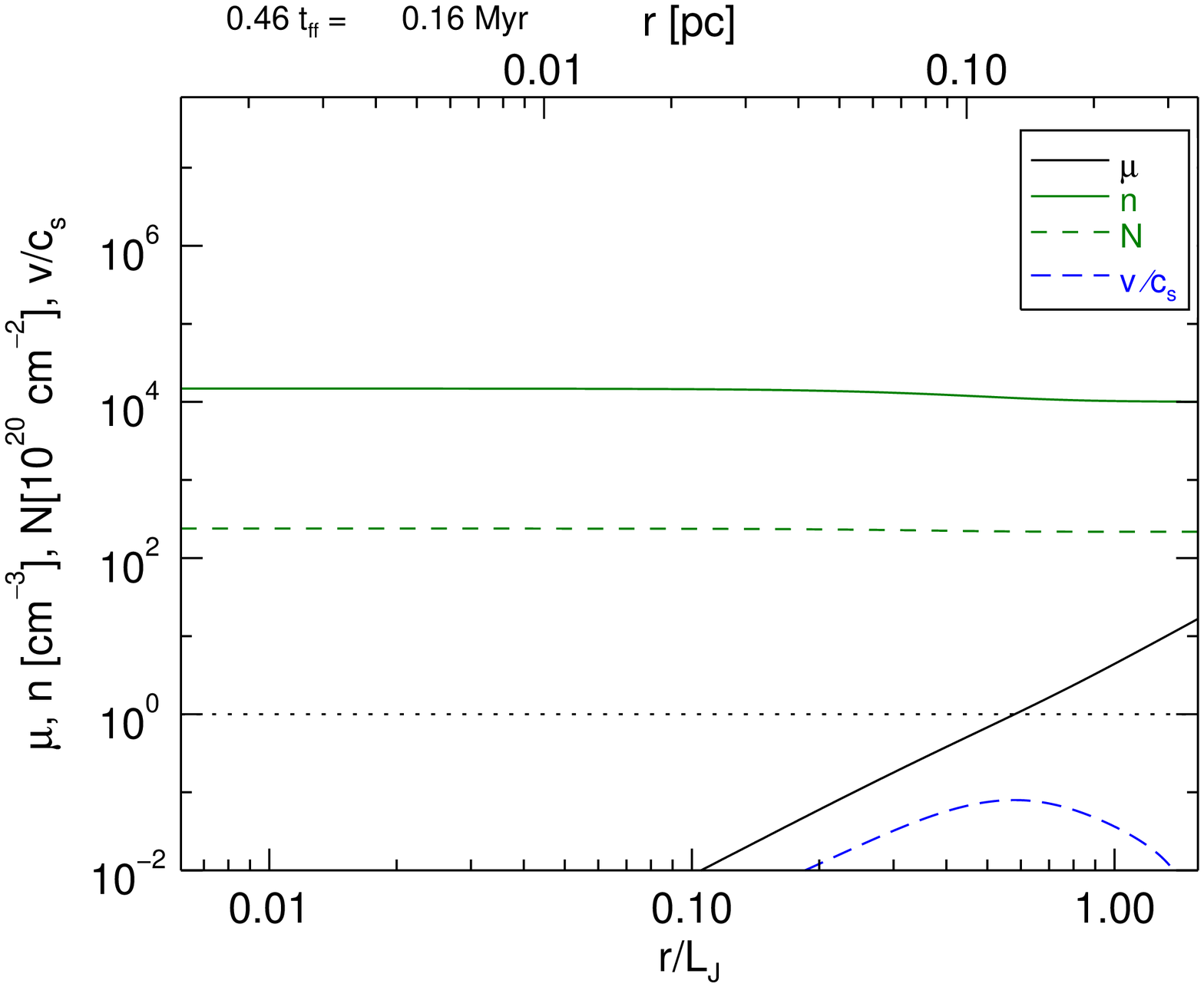}\\
\includegraphics[scale=.38]{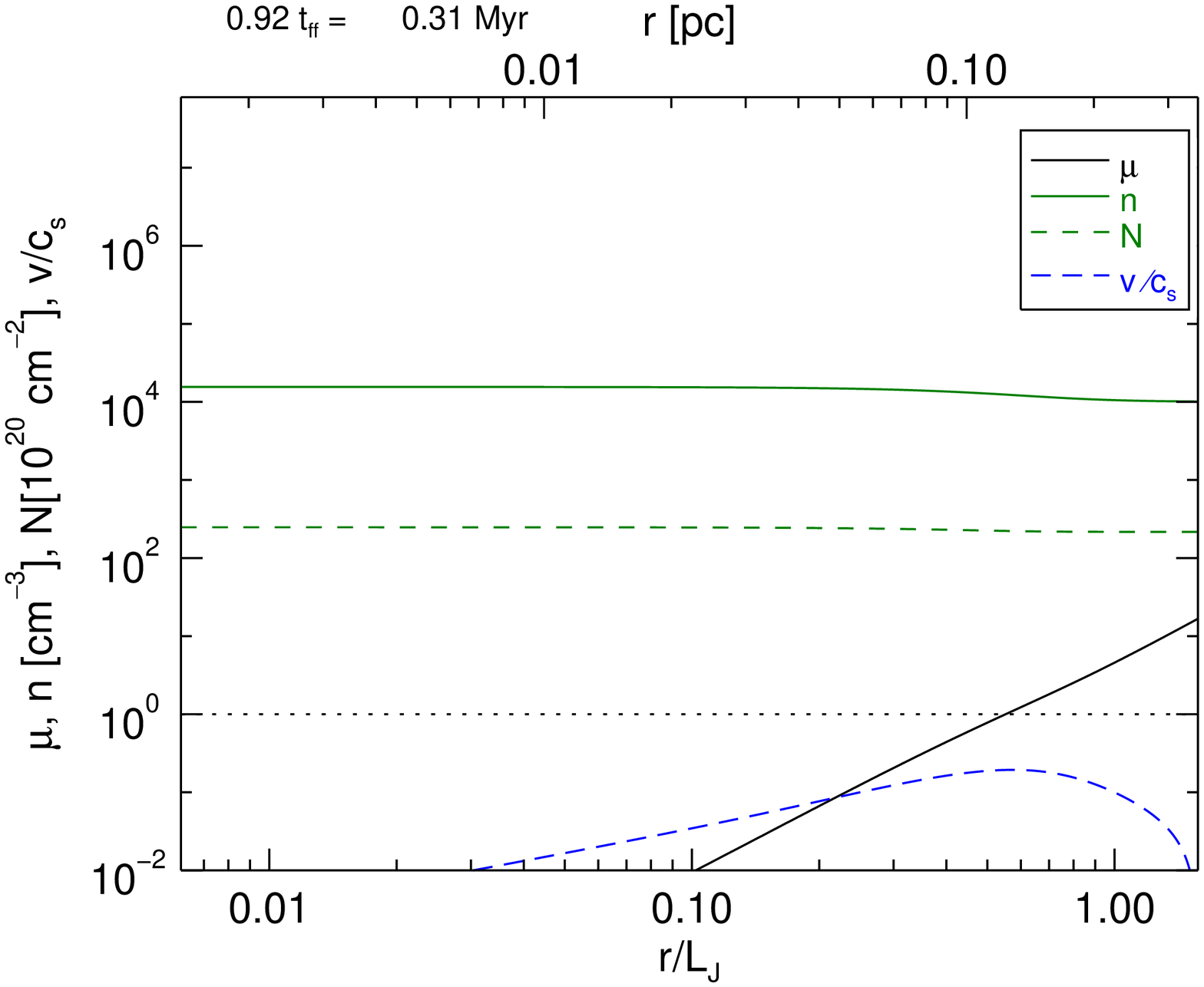}\includegraphics[scale=.38]{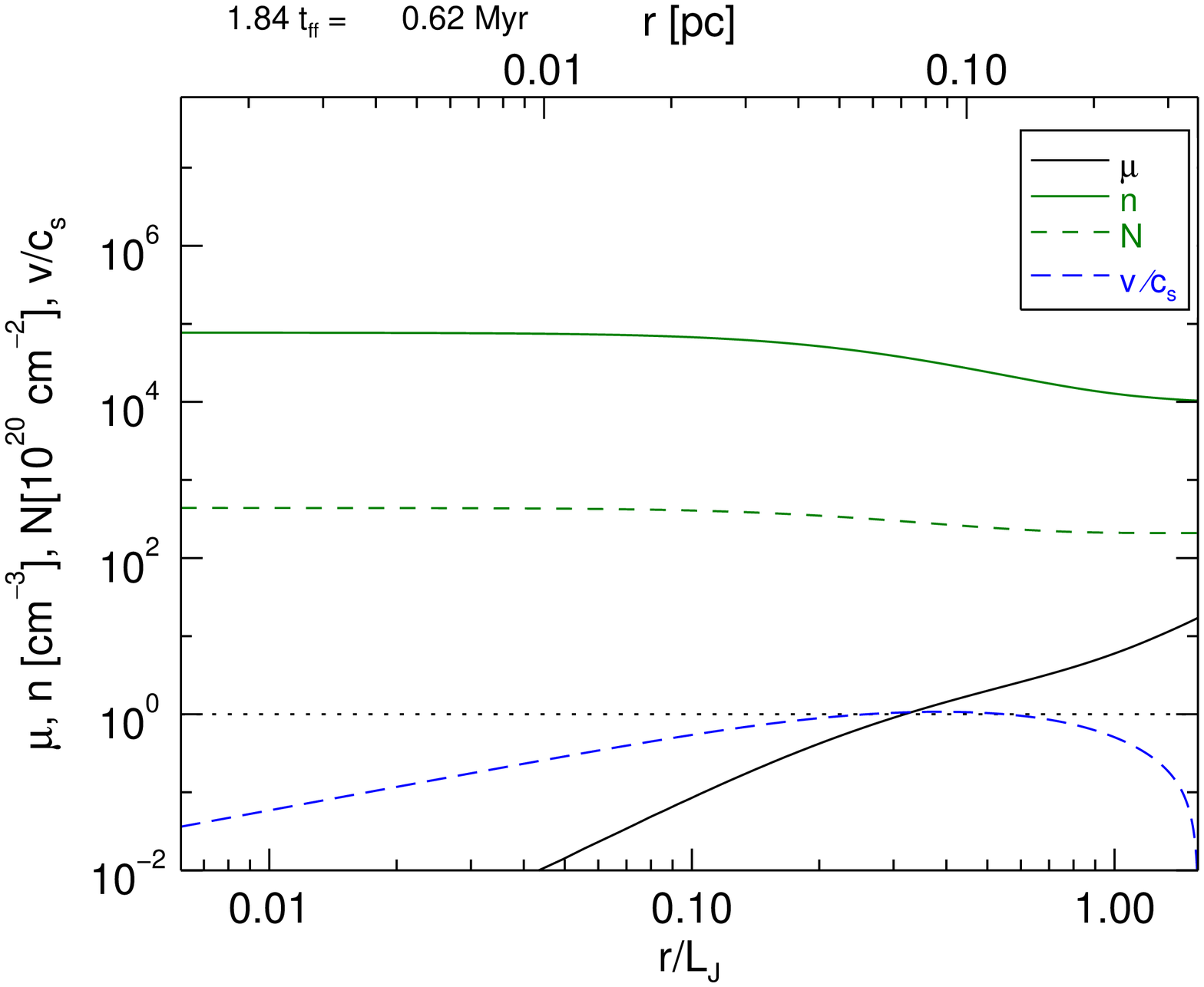}\\
\includegraphics[scale=.38]{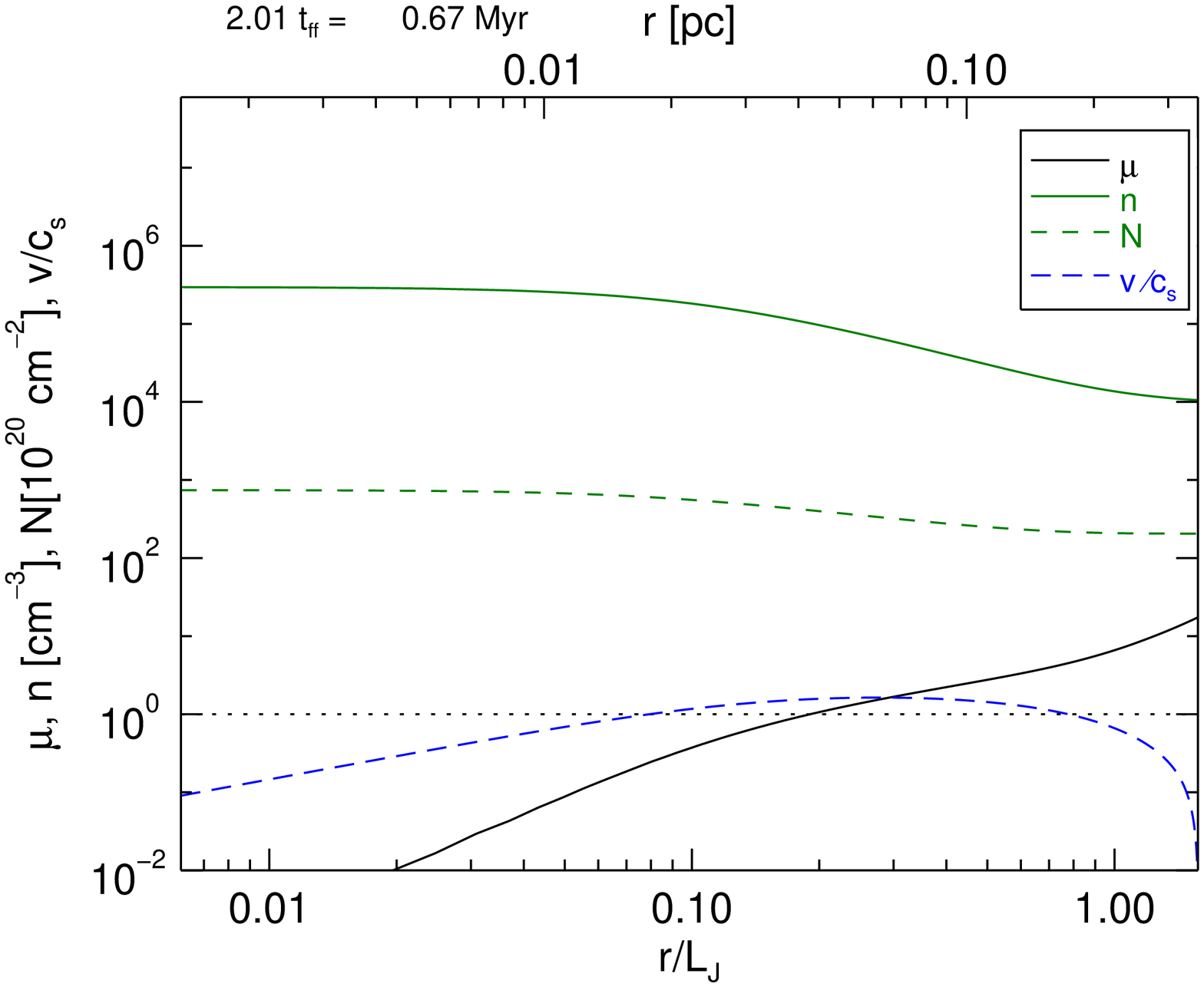}\includegraphics[scale=.38]{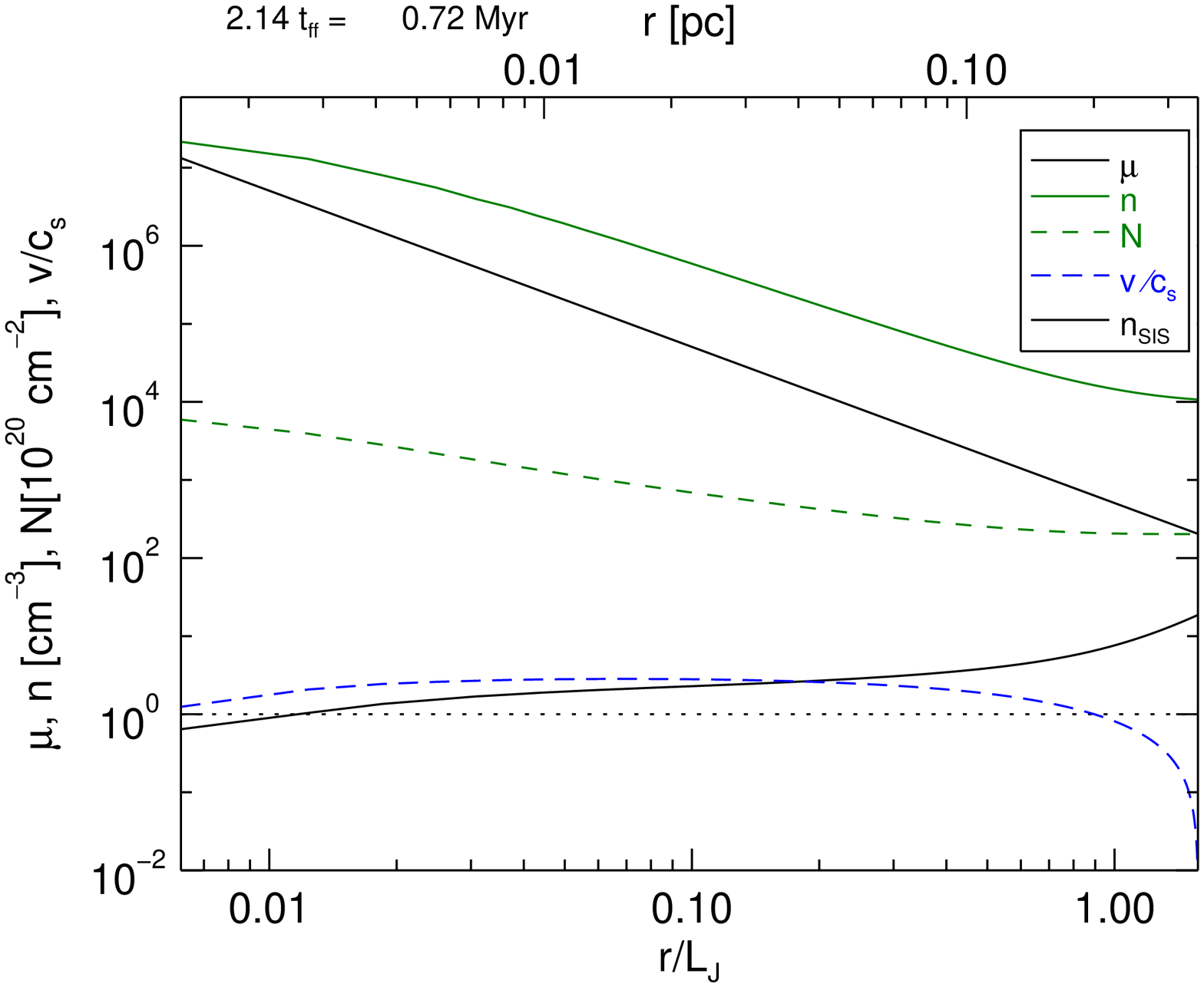}
\caption{$\mu$ ratio, density, column density and velocity
  profiles of the collapsing core (see text for description). 
The panels correspond to snapshots $0$, $14$, $28$, 
$56$, $61$ and $65$, out of a total of $66$ snapshots. The solid black
line in the last panel represents an SIS density profile given by $\left(\cs/2\pi G r^{2} \right)(1/\mu_{\rm w} m_{H})$, where $\mu_{\rm w}$ is the mean particle weight and $m_{\rm H}$ is the mass of hydrogen.\label{fig:all_profiles}}
\end{figure*}

In Fig.\ \ref{fig:all_profiles}, we again show the density and velocity
profiles at six selected snapshots, together with the average column
density and the ratio of the mass to the mean Jeans mass inside each
radial position, $\mu \equiv M(r) /\left\langle M_{J}(r)\right\rangle$.
In this figure, we use a logarithmic radial axis to emphasize the
internal structure of the core. 
The selected snapshots are: $t=0$ (panel {\it a} of Fig.\
\ref{fig:all_profiles}), showing the initial conditions; $t = 0.46\,
\tff$ (panel {\it b}), where $\tff$ is the free-fall time of the
background density. At this time, the gas has developed some moderate
velocity in its external parts, near the {\it Jeans point} (the radius
at which the mass ratio $\mu$ equals unity); $t = 0.92\, \tff$ (panel
{\it c}), in which we begin to see a radially linear subsonic velocity
profile and a uniform density profile within the region delimited by the
Jeans point; $t = 1.84\, \tff$ (panel {\it d}), at which the flow
develops a sonic point ($v=\cs$) at $r \approx 0.19 \LJ \approx 0.05$
pc, located very close to the Jeans point ($\mu=1$) at $r \approx 0.23$
pc; $t = 2.01\, \tff$ (panel {\it e}), at which the transonic point has
already split into two points, bounding a region of almost uniform
supersonic inward velocity (Mach number $\Mach \sim 1.40$); finally, $t
= 2.14\, \tff$ (panel {\it f}), at which the core has an almost uniform
supersonic velocity profile everywhere, with Mach number $\Mach \sim
1.95$, while the density adopts a power-law profile, with a slope
approaching $r^{-2}$. On the last panel, we have overploted an
  SIS density profile (solid black line) for comparison to the actual
  profile of our core, which is clearly more extended than an SIS,
  although with a slope in its envelope that approaches the $r^{-2}$
  slope of the SIS.  Table~\ref{tab:core_parameters} lists various
physical properties of the core at the same six selected snapshots as in
Fig.\ \ref{fig:all_profiles}.

\section{Discussion and implications}
\label{sec:disc}

\subsection{Comparison with observations: $\Mcore/\MBE$ {\it vs.} $\Mcore$ diagram}
\label{sec:comp_obs}

\citet[] [hereafter L08] {Lada08} studied the stability of the dense
core population in the Pipe Nebula region. They defined the cores using
an extinction threshold of $A_{\rm V} = 1.2$ mag, and derived median
values of the density and mass of $n_{H_{2}} = 7.1 \times 10^3 \icmc$,
$M_{\rm core} \sim 0.2$--$20\, \Msun$, respectively, and argued that most
of the cores are gravitationally unbound and thermal-pressure-confined,
with the external pressure provided by the weight of the surrounding
MC. They modeled the cores as BE-spheres, assuming a temperature of
$10\K$, in order to determine their stability, finding that the entire
population is characterized by a single critical BE-mass ($M_{BE}=1.82\left(\langle n\rangle/{10^{4}\icmc}\right)^{-0.5}\left(T/{10\K}\right)^{1.5}\Msun$) of $\sim
2 \Msun$. Most relevant to our interest here is that they plotted the ratio of
the cores' mass to their BE-mass ($\Mcore/\MBE$) {\it vs.} the cores'
mass (see their Fig.\ 9), finding that the observed core sample occupies
a well-defined locus in this diagram.

Subsequently, \citet[] [hereafter R09] {Rathborne09} obtained a more robust
determination of the physical properties of the Pipe cores by combining
extinction and molecular-line data. They found mean radii
$\langle R\rangle\sim0.09\pc$, densities $n_{H_{2}}
\sim 7.3\x10^{3} \pcc$, non-thermal velocity dispersions $\sigma_{\rm
nt}\sim0.18 \kms$, and masses $0.2\, \Msun \leq M_{\rm{core}}\leq 19.4 \Msun$, 
so in what follows we use the R09 sample for our discussion. In addition, 
\citet[] [hereafter I07] {Ikeda07} have carried out a core survey 
in the Orion A molecular cloud. They derived core masses in the range of 
$2$--$80$ $\Msun$, with a mean mass of $12 \pm 12\, \Msun$, a mean density 
$n \sim 2 \times 10^{4}\icms$, consistent with the critical density of 
the H$^{13}$CO$^{+}$(1-0) transition, and mean
velocity dispersions of $0.52 \pm 0.17\kms$. This corresponds to a more
massive and supersonic core sample than the Pipe sample, and thus
offers an interesting complementary set.

In order to compare our simulated core to the observational data, we
must precisely define its boundaries. This is one of the most
challenging tasks when investigating cores, because the boundaries often
depend on circumstantial factors such as the tracer used,
the signal-to-noise ratio, the background level, etc.  In
practice, the core's boundary is often defined by finding gradient
breaks in the column density profiles \citep[e.g., ] [] {Andre+14}---for
example, the radial position of the point where the core appears to
merge with the background, where the profile changes from a power law
(the envelope) to a uniform-density (the background).  Our setup,
embedding the core in a uniform background, naturally lends itself to
this definition. Thus, we operationally define the core's boundary as
the radius at which the density is a certain (small) multiple of the
background density.  Specifically, we consider density thresholds of
$1.125$, $1.25$, and $1.5$ times the background density as the
boundary of the core.

\begin{figure*}[pt]
\centering
\includegraphics[scale=0.36, keepaspectratio]{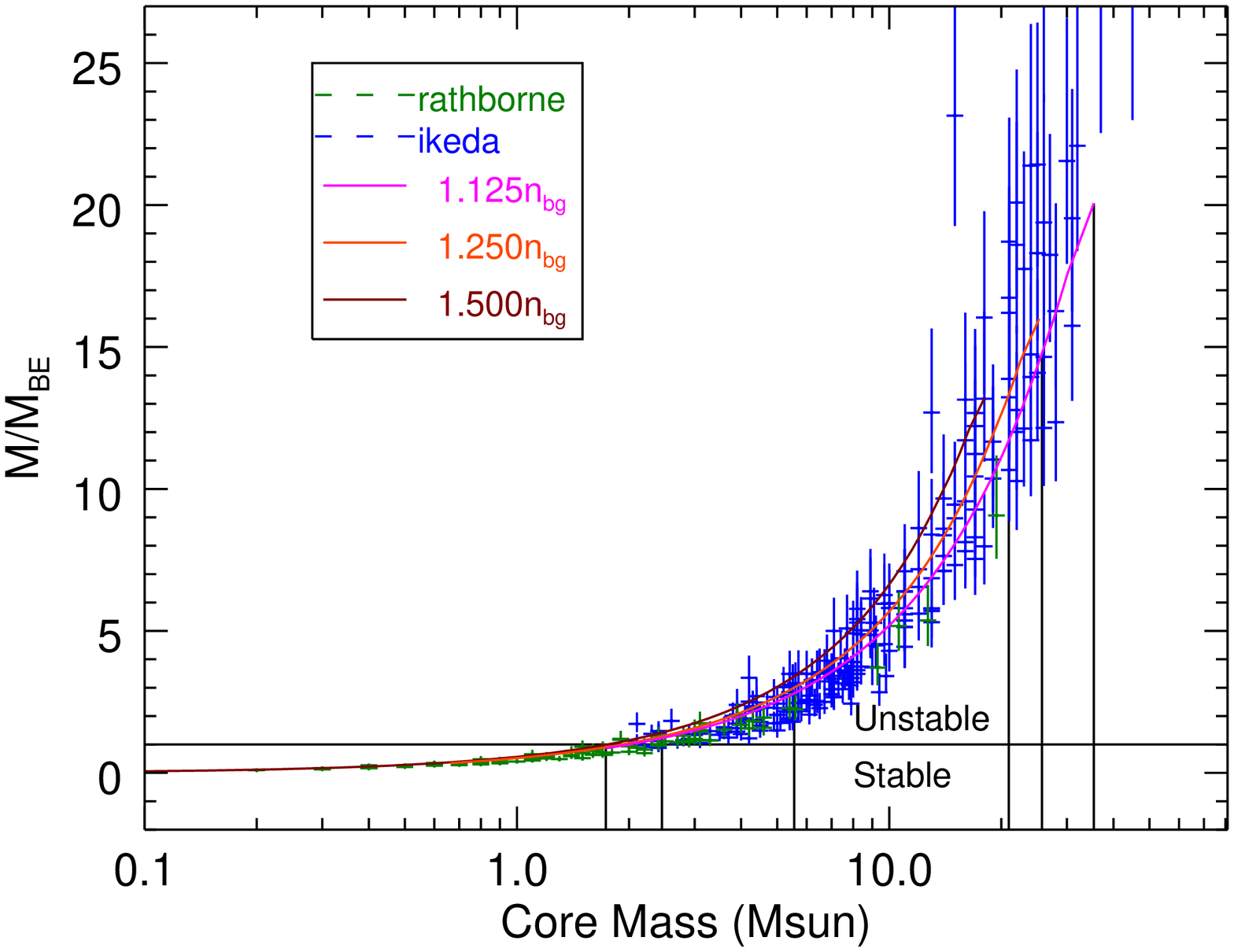}
\includegraphics[scale=0.36, keepaspectratio]{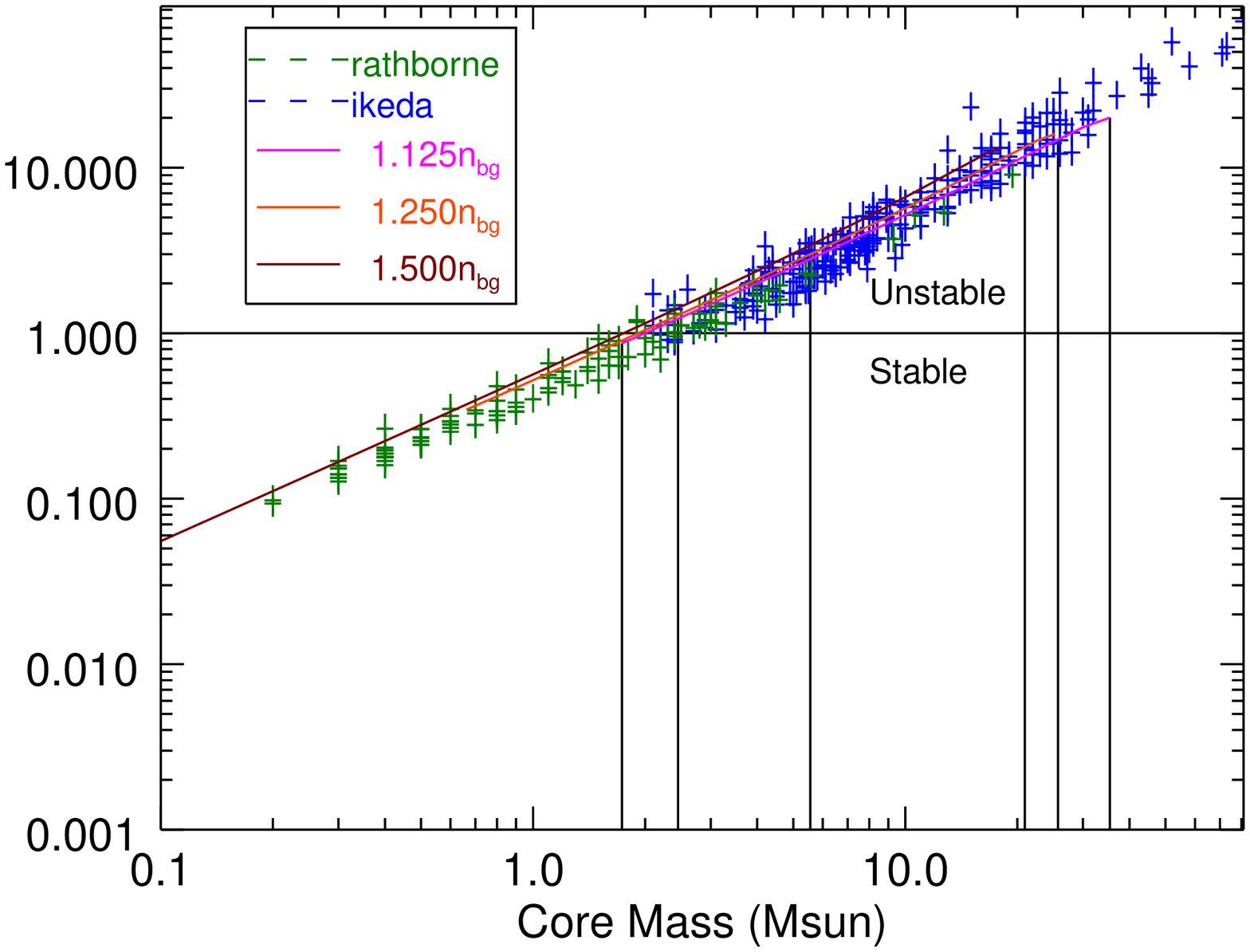}
\caption{\footnotesize Ratio of core mass to BE critical mass
    for the cores in Table 2 from \citet{Rathborne09} (green), Table 1
    from \citet{Ikeda07} (blue) and our simulated core at different
    background density thresholds (continuous curves; see text). Each
    point has a vertical error bar that spans the values of the BE-mass at
    temperatures of $10\K$ and $13\K$ (from top to bottom). The vertical
    lines denote the times shown in Fig.~\ref{fig:all_profiles}.
\label{fig:mass_ratio_evol}}
\end{figure*}

Figure \ref{fig:mass_ratio_evol} shows the ratio $\Mcore/\MBE$
\textit{vs.} $\Mcore$ for the aforementioned observations and for 
our simulated core throughout its evolution, as defined by the three
density thresholds. The times corresponding to the selected
snapshots shown in Fig.\ \ref{fig:all_profiles} are indicated by
vertical lines. For our core, we compute the BE-mass using the
nominal temperature of $11.4 \K$. For the observational core sample, no
explicit temperature information is provided by the authors. R09 assume
a gas temperature of 10 K in all cases, while I07 mention that the
typical temperature in Orion is $20\K$, but make no explicit mention
that this may be the actual temperature in the dense cores. In fact, it
is quite likely that the temperature there is lower, because of the
higher densities. Thus, we have calculated the BE-mass for all the
observed cores assuming $T \approx 11.5\K$, and assigning error bars
whose extremes correspond to $T = 10$ and $13$ K.

From Fig.\ \ref{fig:mass_ratio_evol} we note that the Orion cores
from the I07 sample essentially ocuppy the same locus as the Pipe
cores from the R09 sample in this diagram, although extending towards
higher masses and/or higher values of the mass ratio. Also,
we note that the evolutionary track of our core, as defined by the
threshold at $1.125$ times the background density, tracks almost exactly
the locus of the observed cores. These results strongly suggest that {\it the
sequences of observed (both low- and high-mass) cores are all part of
a self-similar collapse process, only at different evolutionary stages
and total masses, including the apparently stable ones}. The
latter only appear as stable because they are in an early stage of
development, with only a small density contrast over ther background, but
they are nevertheless growing, as their entire background is
gravitationally unstable.

\subsection{Comparison with earlier numerical and analytical collapse studies}
\label{sec:comp_num_analy}

In agreement with previous studies \citep[e.g.,] [MS13] {Larson69,
WhitworthSummers85, VB05, Gomez07}, the density profile
of the simulated prestellar core resembles a BE-sphere at all times,
being flat at the center, and developing a power-law at the external
parts, to which we refer to as ``the envelope''.  As the collapse proceeds,
the central part increases in density while decreasing in radius. 
A supersonic region then emerges, bounded by two transonic
points that become increasingly distant from each other, one moving
inwards and the other outwards, as time
progresses. At the latest stages, just before the formation of the
protostar, the density profile assumes a SIS-like shape, but truncated
at the radius at which the core merges with the background.

These results can be put in the context of early analytical studies on
spherical gravitational collapse that have commonly used similarity
techniques \citep{Larson69, Shu77, WhitworthSummers85}. As is well known,
S77 considered the collapse of a hydrostatic SIS, finding, among others,
the classical ``inside-out'' solution, characterized by a collapsing
inner region bounded by an expanding rarefaction wavefront, beyond which
the gas is static. Inside the transition front, the core is
characterized by density and infall velocity profiles that scale as
$r^{-3/2}$ and $r^{-2}$, respectively. However, a slightly less well
known fact is that S77 discussed, more generally, the large-radius (or
early-time) asymptotic behavior of those solutions for which the
velocity is not initially zero everywhere, but instead only approaches
zero in this limit.  These solutions are characterized by an initial
density profile of the form
\begin{equation}
\rho(r)= \frac{c_{\rm{s}}^{2}A}{4\pi G} \, r^{-2},
\label{eq:shu_density_profile}
\end{equation}
and initial velocity profile given by
\begin{equation}
u(r)= -c_{\rm{s}}^{2} \left(A-2\right) \frac{t}{r},
\label{eq:shu_velocity_profile}
\end{equation} 
where the minus sign indicates that the velocity is directed inwards,
$\cs$ is the isothermal sound speed, and $A$ is a constant that
determines whether the velocity is initially zero and remains at that
value ($A=2$), or instead starts with a finite value and further
increases with time ($A>2$). The former static solution corresponds to a
hydrostatic SIS, while the latter dynamic solution corresponds to an
SIS-like structure (\textit{i.e.} with an $r^{-2}$ density profile) but
with a larger mass and with an initial inward velocity, so that it is
gravitationally unstable and collapses after $t=0$. In this case, the
outer parts of the core are never at rest. It is important to
recall that S77 also calculated the accretion rate onto the protostar
after the formation of the singularity\footnote{In the earlier literature,
the moment at which the singularity forms is referred to as ``core
formation''. Here we avoid this nomenclature in order to avoid
confusion with the dense core of the molecular cloud.} (the protostar
itself) as
\begin{equation}
\dot{M}= \frac{m_{0}c_{\rm{s}}^{3}}{G},
\label{eq:shu_accretion_rate}
\end{equation} 
where $m_{0}$ is a constant related to $A$. The ``canonical'' value,
$m_{0}= 0.975$, corresponds to $A\rightarrow 2^{+}$, \textit{i.e.} to a
hydrostatic initial condition. In addition, S77 argued that the earlier
similarity solution found by LP was unrealistic because its asymptotic
large-radius (or small-time before singularity formation) limit consists
of a uniform inward velocity of $-3.3 c_{\rm{s}}$ and a density that is $4.4$
times that of the SIS at each radius, which he deemed ``unlikely to
occur in a natural way''. However, subsequent analytical and numerical
studies \citep{Hunter77, WhitworthSummers85,
FosterChevalier93} that considered a larger region in parameter space,
showed the existence of a continuum of solutions and, in particular, the
numerical simulations of FC93 showed that the LP solution is approached
only over a finite radial extent, ameliorating S77's objection to it.

Moreover, S77's inside-out solution has been criticized by some authors
\citep[\eg][]{Whitworth+96, VS+05} because its initial condition, the
hydrostatic SIS, is an \textit{unstable} equilibrium, onto which it is
impossible for the core to settle within the context of a turbulent, dynamic
medium such as a molecular cloud. It has also been reported that
observations of molecular cloud cores exhibit extended infall radial
motions that are inconsistent with the inside-out solution of S77
\citep[\eg][]{Tafalla+98, Lee+01}.

Most early numerical simulations considered the case of bounded,
marginally gravitationally unstable initial conditions, neglecting the
possibility of accretion onto the core during the prestellar phase
(prior to the formation of the central singularity). More recently,
accretion has been addressed by various groups \citep[] [MS13] {VB05, Gomez07,
GO09}. \citet{Gomez07} and \citet{GO09} considered the
formation and collapse or re-expansion of a core formed by a
spherically-symmetric compression. They found that cores formed in this
way are bounded by an accretion shock, inside of which the core evolves
along a sequence of BE-configurations, but with a mass that increases
over time due to the accretion until the configuration becomes unstable
and collapses, or else rebounds, if the accretion is insufficient to
render it Jeans-unstable. On the other hand, MS13 considered a constant
accretion flow onto the core, imposed as a boundary condition. Although
their simulations behaved very similarly to ours, the accretion they
used was imposed rather than self-consistent, and forced to be subsonic
at all times.

Very recently, \citet{Keto+15} have investigated the collapse process of
various types of structures. Among other cases, they considered the
collapse of an SIS, of an unstable-equilibrium (UE) BE-sphere\footnote{I.e., 
a solution of the Lane-Emden equation, truncated out to a radius large enough 
that the ratio of central to peripheral density is larger than that of 
a critical BE-sphere, and slightly perturbed so that it proceeds to collapse.}, 
and of a non-equilibrium (NE) pseudo-BE-sphere\footnote{This is not a true BE-sphere as it is
not a solution of the Lane-Emden equation. Instead, it is constructed by
obtaining one such solution and then increasing the density everywhere
by 10\%, so that the configuration is nowhere in equilibrium anymore.}. They found 
that the latter two configurations retained a BE-like density profile, but 
that their velocity profiles differed significantly, with the NE sphere 
developing large velocities out to its truncation radius, while the UE one 
retained a nearly zero velocity at its edge. However, accretion was
not included in their simulations.

Our simulation, instead, includes the novel feature of enveloping the
collapsing core in a globally gravitationally unstable, {\it uniform}
background, to mimic the physical conditions of cores in globally,
hierarchically collapsing clouds, and the fact that cores are often
observed to exist over a roughly uniform background \citep[e.g.,] [and
references therein] {DiFrancesco07, Andre+14}. This situation falls
within the ``band 0'' 
class of solutions investigated by WS85, and essentially exhibits the
behaviour envisaged by those authors: The pre-singularity (prestellar) collapse
proceeds in an outside-in fashion \citep[i.e., at early times the
maximum infall speed occurs at the edge of the core; see also] [] {GO11}
and consists of two main regions: an inner region, characterized by a
roughly flat density profile with an infall velocity that increases
linearly with radius, and an outer region, characterized by an $r^{-2}$
density profile with a uniform infall velocity. However, in our simulation, 
this uniform-infall-speed region is finite and bound
by the outer transonic point, beyond which the infall speed rapidly
drops to zero again.\footnote{In our simulation, the infall motion can
not reach the boundary of the box because of the periodic boundary
conditions, but the boundaries are far enough from the collapsing core
that we do not expect them to significantly affect its evolution during
the time interval we explore.} These results are easy to understand
intuitively. Since the collapse is local, far from its center we should
not expect any motions directed towards it.  Nevertheless, the outer
transonic point moves outwards, implying that the collapse spreads out
to larger regions. This is qualitatively similar to S77's expanding
rarefaction wave although, in our case, the transition from collapsing
to static occurs smoothly, rather than through an abrupt
wavefront. Finally, inside the inner transonic point, the flow exhibits
an infall velocity that is linear with radius, implying that
{\it the velocity smoothly approaches zero towards the center during the
entire prestellar evolution of the core.} (See further discussion of the
implications of this property in Sec.\ \ref{sec:implic_obs_samp}.)

Consequently, during its prestellar stage, our simulation evolves
through a sequence of BE-spheres of increasing central-to-external
density ratios, so that it appears like a \textit{stable} BE-sphere at
early times and like an \textit{unstable} one at later times. The
inner transonic point migrates inwards and reaches the centre at the
time of singularity formation. At this time, the entire core is
characterized by a density configuration that resembles a SIS, except
that in our case this state is \textit{not} hydrostatic, but rather
quite dynamic, with a uniform infall velocity of $\sim 3\, \cs$, thus
corresponding to S77's initial condition with $A>2$.

Our setup differs from studies that start with {\it hydrostatic} BE or
BE-like spheres \citep[e.g.,] [] {FosterChevalier93, VB05, Keto+15}.
Instead, our simulation lets the density and velocity profiles develop
self-consistently, rather than imposing them as initial conditions.
Although it also starts with hydrostatic conditions, these correspond to
an early enough stage that, by the time the core has developed a
pseudo-BE density profile, it is already collapsing everywhere, rather
than being hydrostatic. This can occur, in spite of the very moderate
amplitude of the initial density fluctuation, because the whole
background is unstable, so {\it the fluctuation does not need to exceed the
critical central-to-peripheral density ratio in order to begin
collapsing}. This 
is a reasonable situation since, after all, the core must be {\it
assembled} by moving material from the surroundings into it, and this
requires a non-zero, convergent velocity field.

In order to produce a developed hydrostatic BE-sphere, as in the
standard practice of starting with hydrostatic BE-like structures, the
velocity field that assembled the core would have to first decay to
zero, and then resume again to continue collapsing. This appears as a
highly contrived process. \citet{Keto+15} have suggested that one
process that may temporarily halt the contracting motions is the
turbulent pressure, which must then dissipate to allow the collapse to
resume. However, this implicitly assumes that the non-thermal motions
actually have a sufficiently random nature, and have sufficiently small
characteristic size scales, that they can produce an effective pressure
which provides support against gravity. This notion that has been challenged
recently \citep{VS+08, BP11a, GS+14}.

Conversely, in the hierarchical gravitational collapse scenario, the
possibility of turbulent support is discarded from the outset, and the
motions are assumed to be dominated by an inward component driven by
gravity. Within this scenario, the fragmentation occurs as forseen by
\citet{Hoyle53}, so that, as a cloud contracts and becomes denser, the
average Jeans mass becomes smaller, and so progressively smaller density
fluctuations can collapse. However, because the cloud is turbulent, the
fluctuations are nonlinear, and thus have shorter free-fall times than
the cloud at large, implying that they terminate their collapse
earlier, as observed in numerical simulations of cloud formation and
evolution by converging flows \citep[e.g.] [] {VS+07, VS+09, HH08, Heitsch09,
GV14}. In this scenario, the clumps and cores produced by the turbulence
only act as seeds for local gravitational collapse, and are never supported by 
non-thermal motions, but instead simply grow gravitationally from the
outset, as 
represented in the simulation presented here. As shown in Sec.\
\ref{sec:comp_obs}, this setup naturally explains a
fundamental property of observed dense cores: their location in the
$\Mcore/\MBE$ {\it vs.} $\Mcore$ diagram.

\subsection{Implications for the interpretation of observed core structure}
\label{sec:implic_obs_samp}

In Sec.\ \ref{sec:results} we found that the core evolves along a sequence 
of BE-like density profiles, but with a global infall velocity, so that 
this profile is not indicative of a hydrostatic configuration, not even 
during the apparently stable stages (when the central-to-boundary density 
ratio is smaller than the critical value for instability). This result 
provides a clear interpretation to the ubiquity of BE-like profiles observed 
in prestellar cores without introducing the conundrum that a core needs
to grow in spite of being hydrostatic. The resolution of this dilemma
lies in the 
fact that, although the density profile is BE-{\it like}, it is not a
true BE-sphere in the sense that {\it it does not result as a solution
of the hydrostatic Lane-Emden equation, but rather as a dynamic solution 
of the time-dependent hydrodynamic equations}, which is characterized by 
inwards motion at all times.

Moreover, in Sec.\ \ref{sec:comp_obs} we have shown that our simulated
core, with its boundary defined as the position where the density
becomes equal to the background, traces the locus of the
aforementioned core surveys in the $\Mcore/\MBE$ \textit{vs.}
$\Mcore$ diagram. This result suggests the possibility that those core
ensembles actually represent random samplings of cores at all possible
evolutionary stages, so that their locus in that diagram traces the
evolutionary track of a single core. It also suggests the
possibility that the range of core masses in a given low- or
high-mass star-forming region is determined simply by the mass
increase of the cores during their evolution, but that the initial
fragmentation occurs at the same initial mass, which corresponds
simply to the Jeans mass of the background medium. This is consistent 
with the result by L08 that their whole sample is
well characterized by a {\it single} BE-mass. Indeed, since the
prestellar stages of evolution lead to an approach to an $r^{-2}$
density profile above the background, for which the mean density is
only 3 times larger than the density at the boundary, then the BE
mass, proportional to the inverse square root of the mean density
(assuming isothermality), is at most within a factor $\sqrt{3}$ larger
than that determined by the background. Thus, in the present scenario,
the BE mass is essentially determined by the background density,
rather than by the local core properties. This is also consistent with
the recent finding by \citet{Palau+15} that the number of fragmentats
in massive cores appears to be determined simply by the number of
thermal Jeans masses contained in the core.

Also, we reported that the velocity field is smooth across the core
boundary (compare both panels of Fig.\ 
\ref{fig:dens_vel_evol}). This implies that, if one insists in
describing the core as ``pressure bounded'', the confining pressure is
{\it ram} pressure from the material accreting onto the core, rather
than thermal pressure maintaing a hydrostatic configuration. But
considering it as ``pressure bounded'' is misleading, because any
Lagrangian (i.e., moving with the flow) spherical shell is moving
inwards at any time due to gravity, and thus no bounding is needed. The
correct physical description is that the core is the ``tip of the iceberg'' 
of the globally collapsing cloud.

The fact that the velocity profile in the inner part of the
core is linear with radius during the prestellar evolution implies that
the velocities are smaller closer to the center. This, in turn, implies
that multi-tracer observations of the core would give smaller velocity
dispersions at higher densities/smaller radii, giving the impression of
a ``transition to coherence'' \citep[e.g.,] [] {Goodman98}. However, in 
our scenario, this is not due to dissipation of turbulence, but rather to
the fact that the infall velocities are smaller in the inner part
of the core. Finally, the property that the largest velocities
occur in the outer parts of the core where the density begins to drop,
implies that the line profiles may be narrower than would correspond to
those largest velocities. We dicuss this at more length in Sec.\
\ref{sec:caveats}.

\subsection{Are the supersonic velocities really a problem?}
\label{sec:caveats}

Our simulation, like most other simulations of Larson-Penston-like
flow, develops supersonic velocities in the final stages of evolution,
which however correspond to the stages most likely to be observed, as
the core is most prominent at those times. Such large velocities are
generally not observed in low-mass cores \citep[see, e.g., the review
by] [] {BT07}. MS13 noted this problem, and concluded that their
simulation (quite similar to ours) is not the correct model for the
collapse of actual cores, and that magnetic tension may be necessary to
render the collapse less dynamic. MS13 also pointed out that the
supersonic infall produces accretion rates that are too high, leading to
the so-called ``luminosity problem''
\citep{Kenyon+90}. Our simulation, being spherically symmetric, suffers
from the same problems. However, it is possible that the resolution of
these issues does not lie in invoking the magnetic field to provide
support, but rather in geometrical and/or observational-bias factors.

Concerning geometry, our simulation, like all other non-magnetic,
isothermal, spherically symmetric ones, represents the most dynamic
possible scenario for collapse.  Indeed, recent studies \citep{Toala12,
Pon12} have shown that flattened or filamentary structures collapse on
longer timescales than spherical structures of the same volume
density. Thus, considering the non-spherical nature of the cores may
contribute towards alleviating the problem. Moreover, the luminosity
problem may be resolved if the accretion does not proceed directly onto
the protostar, but rather it is mediated by a circumstellar disk
\citep{KH95, WW01, DV12}, which is also a non-spherical structure. The
resolution of this issue must await analysis of cores arising
self-consistently in fully 3D numerical simulations of molecular cloud
evolution \citep[e.g.,] [] {Smith+13}.

Concerning possible observational biases, we note that observational
determinations of infall velocities depend on the underlying assumptions
for the topology of the velocity
field in the clump. Because a line profile is essentially a
density-weighted radial velocity histogram along the line of sight (with
possible self-absorption features), the fact that in our simulation the
largest velocities occur in the core's envelope rather than at the
center may cause these large velocities to appear 
at the line wings rather than at the central parts of the line, thus
giving the appearance that the infall speeds are smaller than they
actually are. We plan to address this possibility in a future
contribution (Loughnane et al, in prep.).

\section{Summary and Conclusions}
\label{sec:conclusions}

We have presented a highly idealized simulation of the prestellar stages
of the gravitational collapse of an isothermal spherical core within the
scenario of hierarchical gravitational collapse. To accomplish this, we
have embedded the core in a uniform-density background that is, itself,
gravitationally unstable. The evolution of the core was followed since
its earliest stages, starting with a minor density fluctuation of
amplitude 1.5 times the background density, containing a mass slightly
larger than the Jeans mass of the background density, and with a generic
gaussian profile. We have found the following results:

\begin{itemize}

\item The core evolves according to the ``Band 0'' solution of WS85,
which refers to objects that start out far from equilibrium, and
includes the LP solutions.

\item In agreement with previous studies, and with the regularly
  observed structure of prestellar cores, the simulated core develops a
  BE-like profile, with a nearly-uniform-density (or ``flat'') central
  region, and a nearly-power-law envelope, but it is {\it always} in the
  process of collapsing---even during the early stages, when the central
  to background density contrast is smaller than that of a critical BE
  sphere ($\rhoc/\rhob \sim 14$), and the core would be labeled as
  ``stable''. This result removes the apparent inconsistency between the
  apparently hydrostatic density structure of the cores (i.e., of stable
  BE-spheres) and the need for them to grow in order to eventually form
  stars. It also suggests that a smaller fraction of cores are
  ``starless'' (in the sense that they will never form stars) than
  usually thought. Instead, it is possible that these are just in their
  very earliest stages of growth.

\item The collapse proceeds in an ``outside-in'' fashion, developing the
largest (and nearly radially constant) speeds in the power-law density
envelope, and a linear-with-radius velocity profile in the central flat
region. This implies that the velocities in the centermost parts of the
core are small and subsonic during the whole prestellar evolution. This is
contrary to the famous inside-out solution of \citet{Shu77}, which is
{\it not} expected because its initial condition, a {\it hydrostatic}
singular isothermal sphere, is an unstable equilibrium, and therefore
unrealizable from a dynamical previous evolution. 

\item The boundary of the core, defined as the position where it merges
  with the background, increases in radius as time progresses, so that
  the core thus defined effectively grows in mass and size. The velocity
  field is smooth across this boundary, so that the material outside the
  boundary accretes onto the core smoothly, although it does not
  increase its density until it crosses the boundary.

\item The largest velocities in the system---those appearing in the
  power-law envelope---are supersonic by the time the core has
    grown enough to be clearly detectable (with at least a tenfold
    enhancement over the background). However, the fact that the largest
    velocities are located at the envelope implies that they will
    receive a lower density weighting for the production of a line
    profile, so that infall profiles may underestimate the velocities.
    In a future study we will investigate the nature of the lines
    produced by our core (Loughnane et al., in prep.).

\item The ratio of the core's mass ($\Mcore$) to the critical BE-mass
  ($\MBE$) increases as the core's mass increases, and traces the locus
  of observed cores in the $\Mcore/\MBE$ {\it vs.}  $\Mcore$ diagram,
  evolving from apparently stable to apparently unstable configurations,
  although the core is unstable at all times.  Thus, the locus of
  observed cores in this diagram can be interpreted as a random sample
  of evolving collapsing cores within globally unstable clouds. This 
  result suggests that the sequences of observed (both low- and 
  high-mass) cores are all part of a self-similar collapse process, only at 
  different evolutionary stages and total masses, including the apparently 
  stable ones.

\end{itemize}

We conclude that the evolution of marginally unstable cores embedded in
a strongly unstable environment, as prescribed by the hierarchical
gravitational collapse scenario, is not only consistent with the
observed density structure of molecular cloud cores and their location
in the $\Mcore/\MBE$ {\it vs.} $\Mcore$ diagram, but
provides a natural explanation to the problem of how a core may grow in
mass while appearing gravitationally stable.
Another conclusion is that the choice of initial conditions is crucial
for the subsequent evolution of the system, and thus care must be
exercised in choosing the most realistic initial conditions within the
realm of the physical problem at hand.

Some problems remain, of course, regarding the appearance of supersonic
speeds during the prestellar evolution, which are generally not observed
in low-mass cores, as pointed out by MS13. In future contributions, we
plan to investigate whether this apparent discrepancy can be resolved in
terms of a non-spherical collapse geometry and/or the effects of the
outside-in velocity field on the formation of infall line profiles.

\acknowledgements
We thankfully acknowledge an anoymous referee, whose insightful and sharp
recommendations helped in improving the clarity and coherence of the
paper. The numerical simulations were performed in the cluster acquired
with CONACYT grant 102488 to E.V.-S. Also, R.M.L.\ was supported with
funds from this grant.

\end{document}